\documentclass[a4paper,11pt]{article}
\usepackage{pos}
\usepackage{enumitem}

\title{Exploring a PMT+SiPM Hybrid Optical Module for Next Generation Neutrino Telescopes}

\author*[a]{Fan Hu}
\author[a,b]{Zhuo Li}
\author[c,d]{Donglian Xu}

\affiliation[a]{Department of Astronomy, Peking University, 5 Yiheyuan Rd, Beijing, China}

\affiliation[b]{The Kavli Institute for Astronomy and Astrophysics, Peking University, 5 Yiheyuan Rd, Beijing, China}

\affiliation[c]{Tsung-Dao Lee Institute, Shanghai Jiao Tong University, 800 Dongchuan Rd, Shanghai, China}

\affiliation[d]{School of Physics and Astronomy, Shanghai Jiao Tong University, 800 Dongchuan Rd, Shanghai, China}

\emailAdd{fan\_hu@pke.edu.cn}
\emailAdd{zhuo.li@pku.edu.cn}
\emailAdd{donglianxu@sjtu.edu.cn}

\abstract{Cosmic neutrinos are unique probes of the high energy universe. IceCube has discovered a diffuse astrophysical neutrino flux since 2013, but their origin remains elusive. The potential sources could include, for example, active galactic nuclei, gamma-ray bursts and star burst galaxies. To resolve those scenarios, higher statistics and better angular resolution of astrophysical neutrinos are needed. An optical module with larger photon collection area and more precise timing resolution in a next generation neutrino telescope could help. Silicon photon multipliers (SiPMs), with high quantum efficiency and fast responding time, combining with traditional PMTs, could boost photon detection efficiency and pointing capability. We will present a study on exploring the benefits of combining multiple PMTs and SiPMs in an optical module.}

\FullConference{%
  *** 37th International Cosmic Ray Conference (ICRC2021), ***\\
  *** 12-23 July 2021 ***\\
  *** Berlin, Germany - Online ***
}

\begin{document}

\maketitle

\clearpage

\section{Introduction}

The diffused astrophysical neutrino flux discovered by IceCube has painted the road for the development of the next generation neutrino telescopes: such an isotropic flux could come from a high degeneracy of astrophysical sources such as active galactic nuclei, gamma-ray bursts and star forming galaxies \cite{Astroph2020:2019ows}, to resolve which requires a neutrino telescope with much improved pointing capability \cite{Fang:2016hop}.

Cherenkov neutrino telescopes such as IceCube, KM3NeT and Baikal-GVD probe cosmic neutrinos by detecting the faint light emitted by Cherenkov processes of secondaries produced in neutrino-matter interaction. Those neutrino telescopes are composed of arrays of Digital Optical Module (DOM) that contain one or more Photomultiplier Tubes (PMTs) which can detect low-light signals down to the single-photon level. The pointing of neutrino telescopes is based on the reconstruction of photon hit time at DOMs. The intrinsic transient time spread (TTS), typically of a few nanoseconds, of the PMTs in the DOMs smears the precision of time measurement, which could compromise the angular resolution of neutrino telescopes.

The Silicon Photomultiplier (SiPM) is a modern solid-state sensor that works at the same province of the PMT. The photon electrons produced inside a SiPM are amplified and produce a cascade in-situ, which enables the SiPM to obtain high photon detection efficiency (PDE) and a fast response time down to $O(10)$ ps \cite{Acerbi:2019qgp}. Combining multiple PMTs and SiPMs in one hybrid DOM (hDOM) can take advantage of both the precise photon hit time feature as well as maintaining large photon detection area. Applying such hDOMs in a next generation neutrino telescope could potentially improve its photon detection efficiency and pointing capability.

\begin{figure}[htbp]
    \centering
    \includegraphics[width=0.5\linewidth]{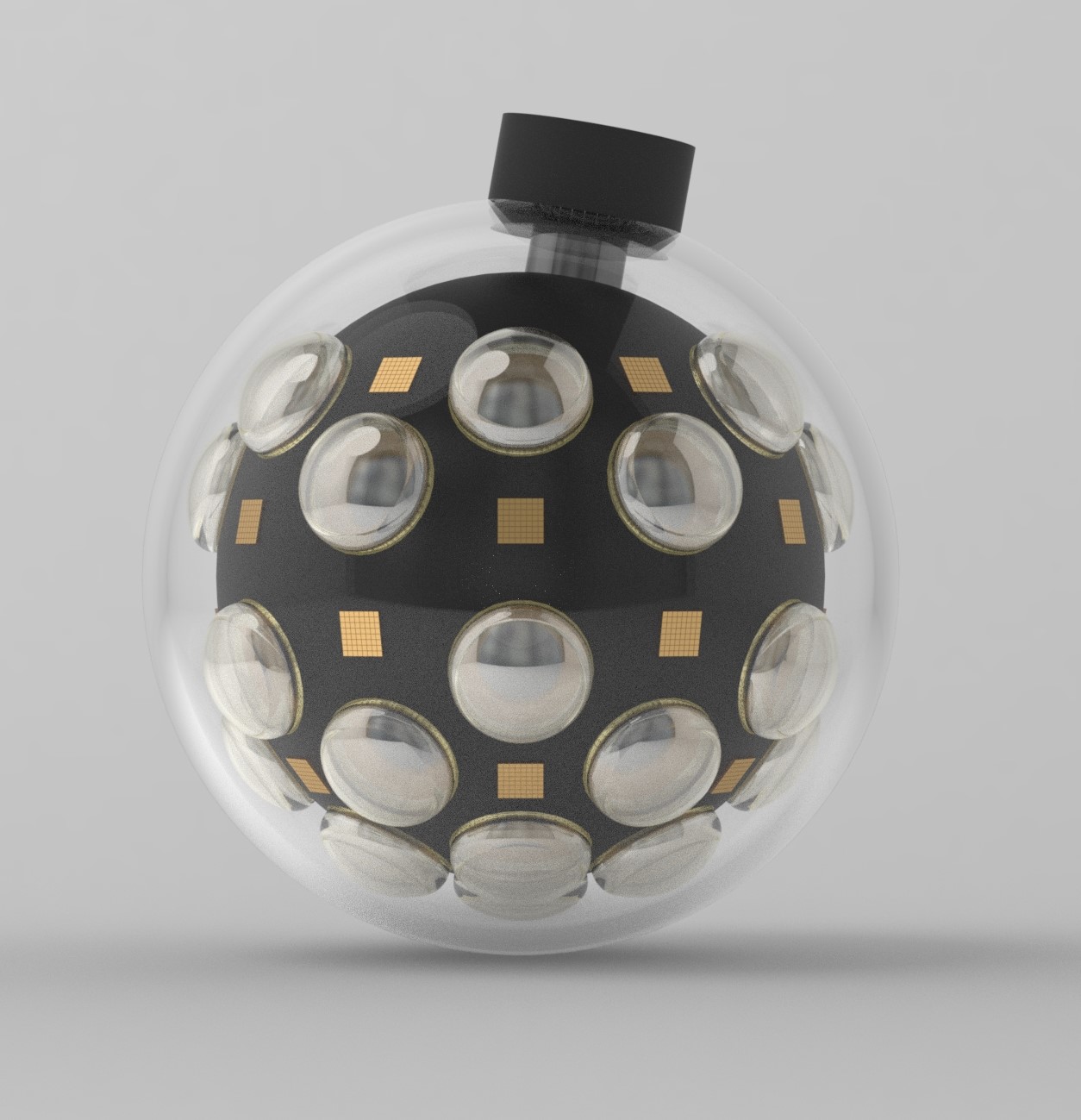}
    \caption{\textbf{Conceptual design of an hDOM that consists 31 PMTs and 20 SiPM arrays.}}
    \label{fig: hDOM Design}
\end{figure}

One way to construct such an hDOM is to add SiPMs among the intervals of PMTs, as shown in figure \ref{fig: hDOM Design}. This conceptual design of an hDOM contains 31 PMTs that distributed the same as the mDOM \cite{KM3NeT:2015huz} produced by KM3NeT to achieve comprehensive comparison. There are additional 20 SiPM arrays distributed in between of PMTs, forming 4 rings along the equatorial direction. 

In the following sections, we will analyze the influence of fast hit time response obtained by SiPMs to the angular resolution for minimally ionizing muons of the neutrino telescope placed in both sea water and Antarctic glacial ice medium in a detailed simulation study.

\section{Simulation Setup}

Our simulation is set up in the Geant4 framework \cite{Geant4:2006ve}. The neutrino telescope is $1~\mathrm{km}^3$ in size, containing $5000$ DOMs. These DOMs are separated by  $100~\mathrm{m}$, $100~\mathrm{m}$ and $20~\mathrm{m}$ in $x$, $y$ and $z$ directions, forming a rectangular grid. There are three different configures of DOMs in our simulation for comparison:

\begin{enumerate}[label={\alph*)}]
    \item PMT DOM that contains only PMT part of our hDOM.
    \item Toy SiPM DOM that replace the PMTs in the first PMT DOM to "toy SiPMs".
    \item hDOM that contains both PMTs and real SiPM arrays as shown in figure \ref{fig: hDOM Design}.
\end{enumerate}

The first PMT DOM is used as a benchmark as it has a similar configuration as the mDOM used in KM3NeT. The second DOM is composed of "toy SiPMs". The "toy SiPM" is the same as the PMT in shape and photon detection efficiency (PDE), but has a smaller transient time spread (TTS). This configuration is used to demonstrate the influence of timing ability on the angular resolution of a neutrino telescope. The last hDOM configuration is used to guide the real production. 

\begin{figure}
    \centering
    \includegraphics[width=1.0\linewidth]{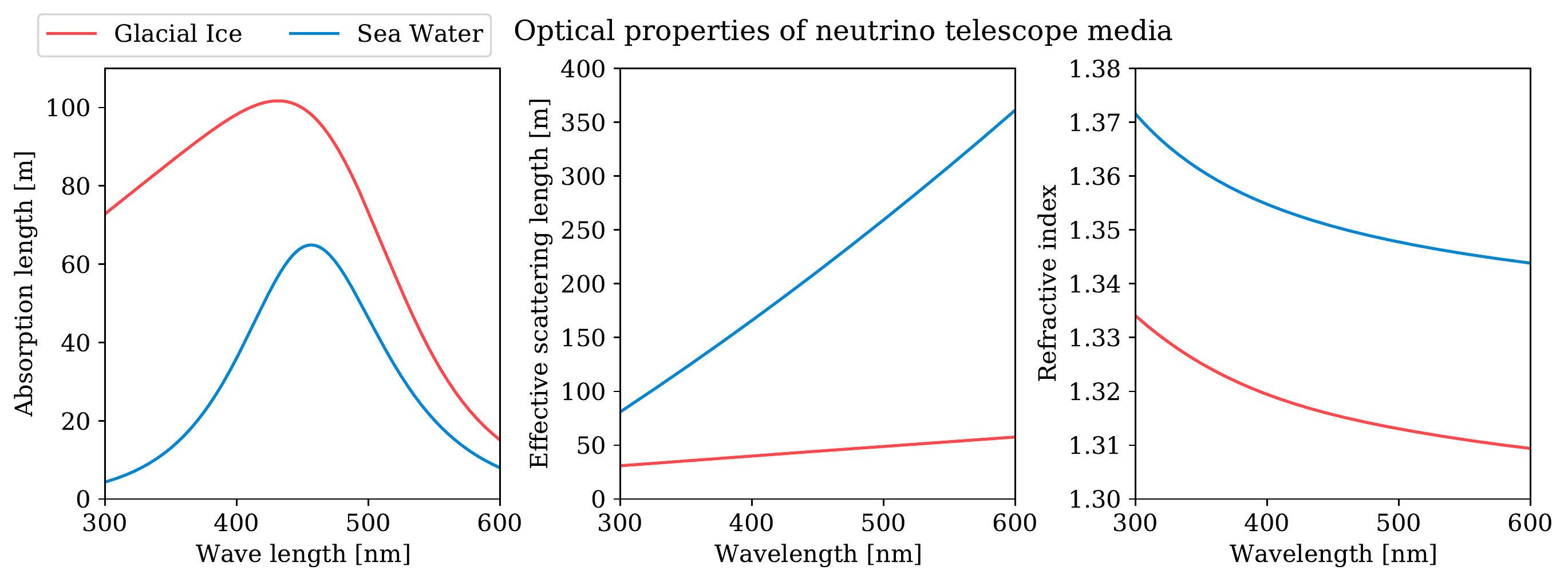}
    \caption{\textbf{Optical properties of water and ice.} Blue curve represents for sea water while red curve represents for Antarctic glacial ice.}
    \label{fig: optical properties}
\end{figure}

In our simulation, the PMT model used is the XP72B20 produced by HZC Photonics, and the SiPM model is PA3325-WB-0808 produced by KETEK. Their main properties are summarized in table \ref{tab: sensor parameters}. The optical property of sea water is referred to ANTARES partic-0.0075 model \cite{Kopper:2010kia}. And we refer IceCube measured data \cite{IceCube:2013rt} for ice optical properties. Those optical properties are shown in figure \ref{fig: optical properties}. The scattering of sea water is composed of 17\% Rayleigh scatter and 83\% Mie scatter with the mean forward ratio of 0.93. And the scattering of glacial ice is Mie scatter only with the mean forward ratio of 0.94.

\begin{table} [htbp]
    \centering
    \begin{tabular}{c|c|c|c}
        \hline
        Sensors & PMT & toy SiPM & SiPM  \\
        \hline
        Physical size [$\rm{cm}^2$] & 40.7  & 40.7 & 7.3 \\
        \hline
        PDE@405 nm [\%] & 24 & 24 & 42 \\
        \hline
        TTS [$\rm{ns}$] & 5 & 0.1 & 0.1 \\
        \hline
    \end{tabular}
    \caption{\textbf{Main parameters of photon sensors.}}
    \label{tab: sensor parameters}
\end{table}

The primary particles in the simulation are $1 ~\rm{TeV}$ muons. They are injected 200 meters away in all directions and propagated towards the inner part of the neutrino telescope. Such a low energy muon lays in the minimum ionization regime and is not likely to undergo stochastic energy loss processes. During its propagation, it produces secondary particles and emits Cherenkov photons along with charged secondary particles.

Photon hit time is defined as the time of photon arriving at the surface of the sensor smeared by a Gaussian distribution that represents the TTS of the sensor. Electronic effects or background noise are not yet treated in the current study but will be taken into account in the future.

An additional simulation is implemented to study the effective photon detection area $A_\mathrm{eff}$ of different components of hDOM as a function of photon incident angle. In the simulation, we uniformly emit photons from a disk source toward the hDOM. The effective area of hDOM is an important parameter to describe the photon detection ability of hDOM and is defined by the following equation:

\begin{equation}
    A_\mathrm{eff} = A_\mathrm{emit} \times \frac{N_\mathrm{receive}}{N_\mathrm{emit}}
    \label{eq: eff area}
\end{equation}

Here $A_\mathrm{emit}$ is the photon emitting disk size in simulation, $N_\mathrm{emit}$ is the number of photons emitted from the source, and $N_\mathrm{receive}$ is the number of photons received by the sensors.

\section{Reconstruction Method}

To study the angular resolution of the neutrino telescope for muon track events, a reconstruction pipeline is built. The reconstruction method used is based on the photon residual times $t_\mathrm{res}$ that are also widely used in previous neutrino telescopes \cite{Trovato:2014msl, Wiebusch:2003}. $t_\mathrm{res}$ is defined as the time difference between the hit time recorded by the sensor and the geometrical time that can be calculated at a given assumption of track parameters. $t_\mathrm{res}$ is expected to be zero if the track parameters are precisely the same as the simulation truth, photon encounters no scattering, and the sensor has no time spread. In the real case, the probability density distribution of residual time $p(t_\mathrm{res} | d)$ can spread largely and has different shapes according to distance.

The reconstruction processes are divided into four steps. First, a least-squares method is used to fit the track parameters to the hits point received in DOMs. Then, an m-estimator is used in maximum likelihood analysis, which takes previous result as initial parameters and gets a better reconstruction accuracy. After that, hits with large residual times are cleaned away based on the track parameters from the previous step. At last, we deploy a multi-photon-electron (MPE) maximum likelihood method that utilizes first hit time information to obtain the final result.

The performance of the reconstruction result depends highly on the accurate description of the probability density function (PDF) of the residual time in the MPE likelihood method. Here, we build a new PDF based on the skew Cathy distribution for the sea water medium as shown in equation \ref{eq: PDF}.

\begin{equation}
    p(\tau) = \frac{1}{\pi (\tau^2+1)} \times \frac{2}{e^{-\alpha \tau} + 1}
    \label{eq: PDF}
\end{equation}

In the equation, $\tau = (t_\mathrm{res} - \mu) / \sigma$ is the shifted and scaled residual time and $\alpha$ is the skew factor. The first term of Eq. \ref{eq: PDF} represents the Cathy distribution and the second term is a sigmoid function that brings in skewness to the PDF. This PDF can capture the feature that many photons are arrived directly at the DOM without any scatter, forming a peak near $t_\mathrm{res} = 0$. It can also cover the hits that arrive late due to deflection of scattering and hits that has $t_\mathrm{res} < 0$ due to sensors' timing error or dispersion effect. The PDF in such form is numerically simple to compute, positively defined in all domains, and insensitive to the late arrived photons. 

\begin{figure}
    \centering
    \includegraphics[width=0.8\linewidth]{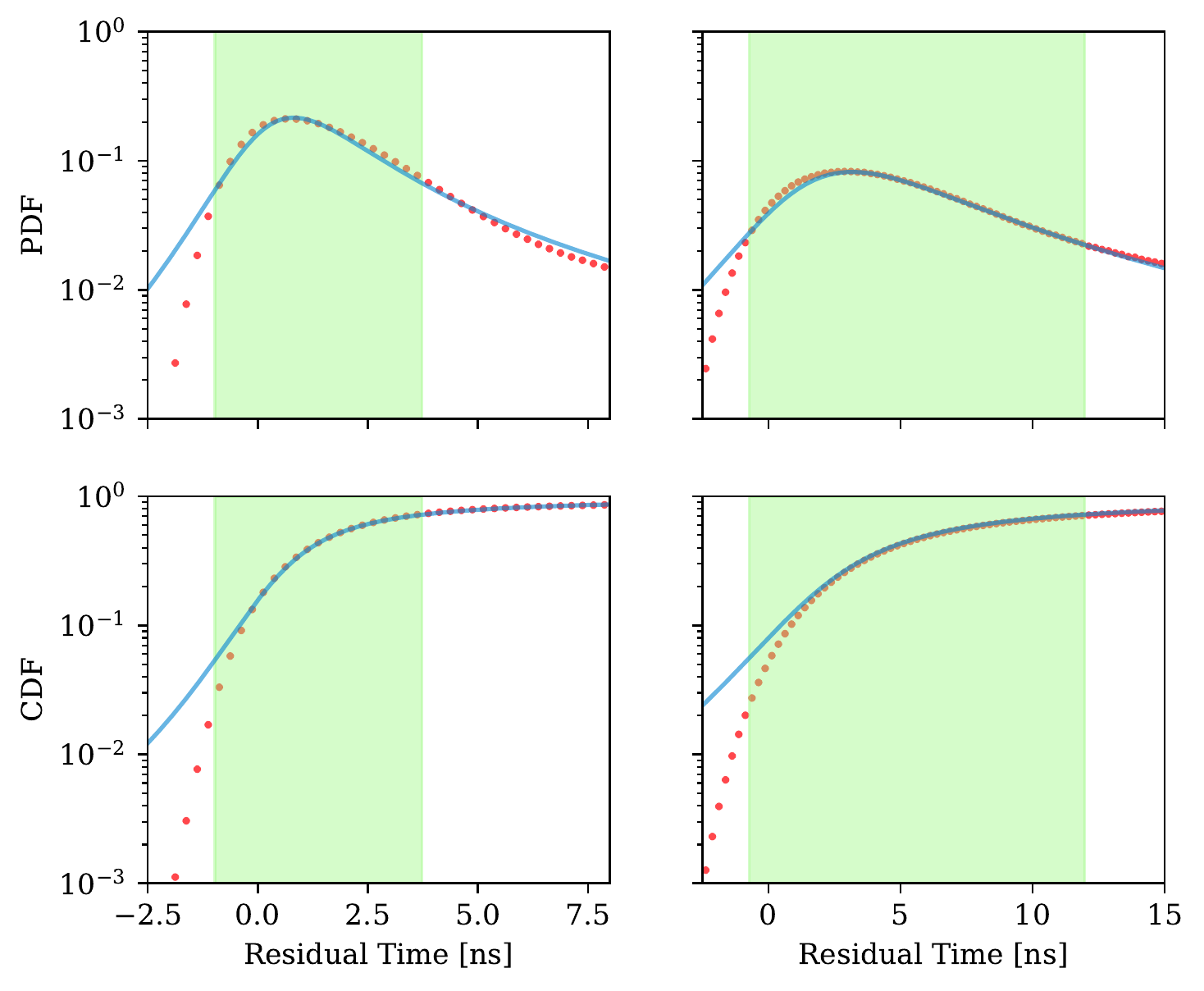}
    \caption{\textbf{Simulated data distribution of the residual time and analytical fit result.} \textbf{Left:} DOM-to-track distance $d = 20 ~\mathrm{m}$, \textbf{Right:} $d = 70 ~\mathrm{m}$. \textbf{Top:} probability density function (PDF), \textbf{Bottom:} cumulative distribution function (CDF). The green bands show the time interval where CDF has value of (2.5\%, 70.5\%). Only data points inside green bands are used in the fit.}
    \label{fig: PDF}
\end{figure}

We fit $\mu$, $\sigma$, and $\alpha$ parameters as a function of DOM-to-track distance $d$. The comparison between fit and data are shown in figure \ref{fig: PDF} and the parameters used are summarized in Eq. \ref{eq: parameters}. Only the hits with residual time lays in the interval (2.5\%, 70.5\%) are used to fit the curve, because those hits are most likely to be the first hit in the DOM. 

\begin{equation}
    \mu = 0.0199 d - 0.114,~~
    \sigma = 0.0669 d + 0.328,~~
    \alpha = 0.0127 d + 0.538
    \label{eq: parameters}
\end{equation}

For the ice medium, the PDF of residual time can not be fit by equation \ref{eq: PDF}. So a 2D cubic spline method is used to get PDF by interpolating the simulated data. It should be noted that such an interpolated PDF is not as robust as the manually constructed PDF in the minimum likelihood reconstruction for the moment.

\section{Results}

\subsection{Effective Area of hDOM}

\begin{figure} [htbp]
    \centering
    \includegraphics[width=1.0\linewidth]{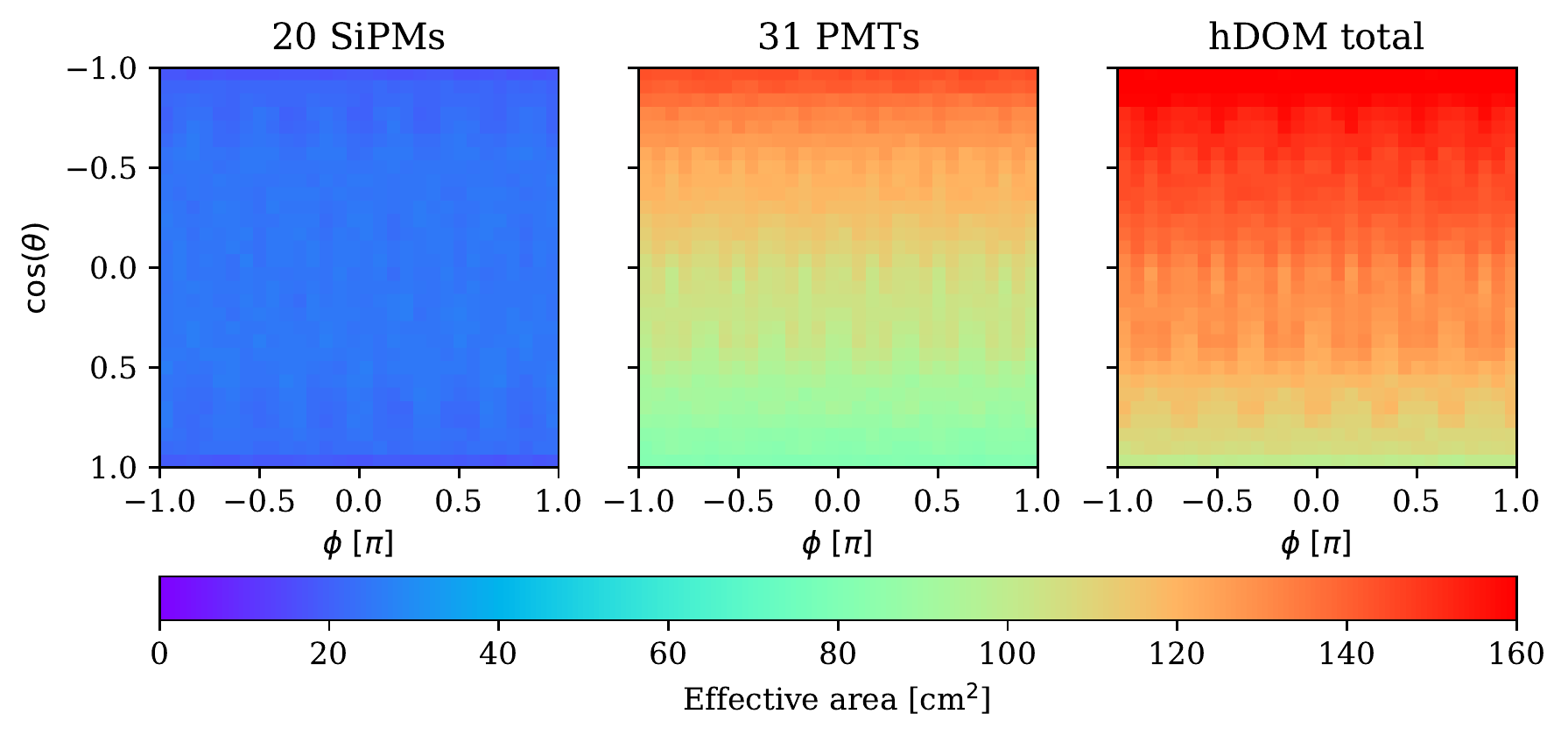}
    \caption{\textbf{Effective area of hDOM.} \textbf{Left:} Effective area of SiPM part in hDOM. \textbf{Middle:} Effective area of SiPM part in hDOM. \textbf{Right:} The total effective area of hDOM.}
    \label{fig: effective area}
\end{figure}

The effective area $A_\mathrm{eff}$ of each photon sensor components of the hDOM is shown in figure \ref{fig: effective area}. $A_\mathrm{eff}$ can be a function of zenith angle due to the distribution of senor in hDOM surface. In our configuration, the sensors' total physical area is about $1260 ~\mathrm{cm}^2$ for PMTs and $146 ~\mathrm{cm}^2$ for SiPMs. As can be seen in the figure, $A_\mathrm{eff, PMT}$ can range from $80~\mathrm{cm}^2$ from north pole to $145~\mathrm{cm}^2$ from south pole. $A_\mathrm{eff, SiPM}$ is about $25~\mathrm{cm}^2$ at equatorial plane to $18~\mathrm{cm}^2$ at pole and is slightly more homogeneous in directions than $A_\mathrm{eff, PMT}$. The overall $A_\mathrm{eff, PMT} : A_\mathrm{eff, SiPM} = 4.6 : 1 $, meaning that about one of 5.6 hits received in hDOM will be SiPM hit.

\subsection{Angular Resolution of hDOM array}

\begin{figure}
    \centering
    \includegraphics[width=1.0\linewidth]{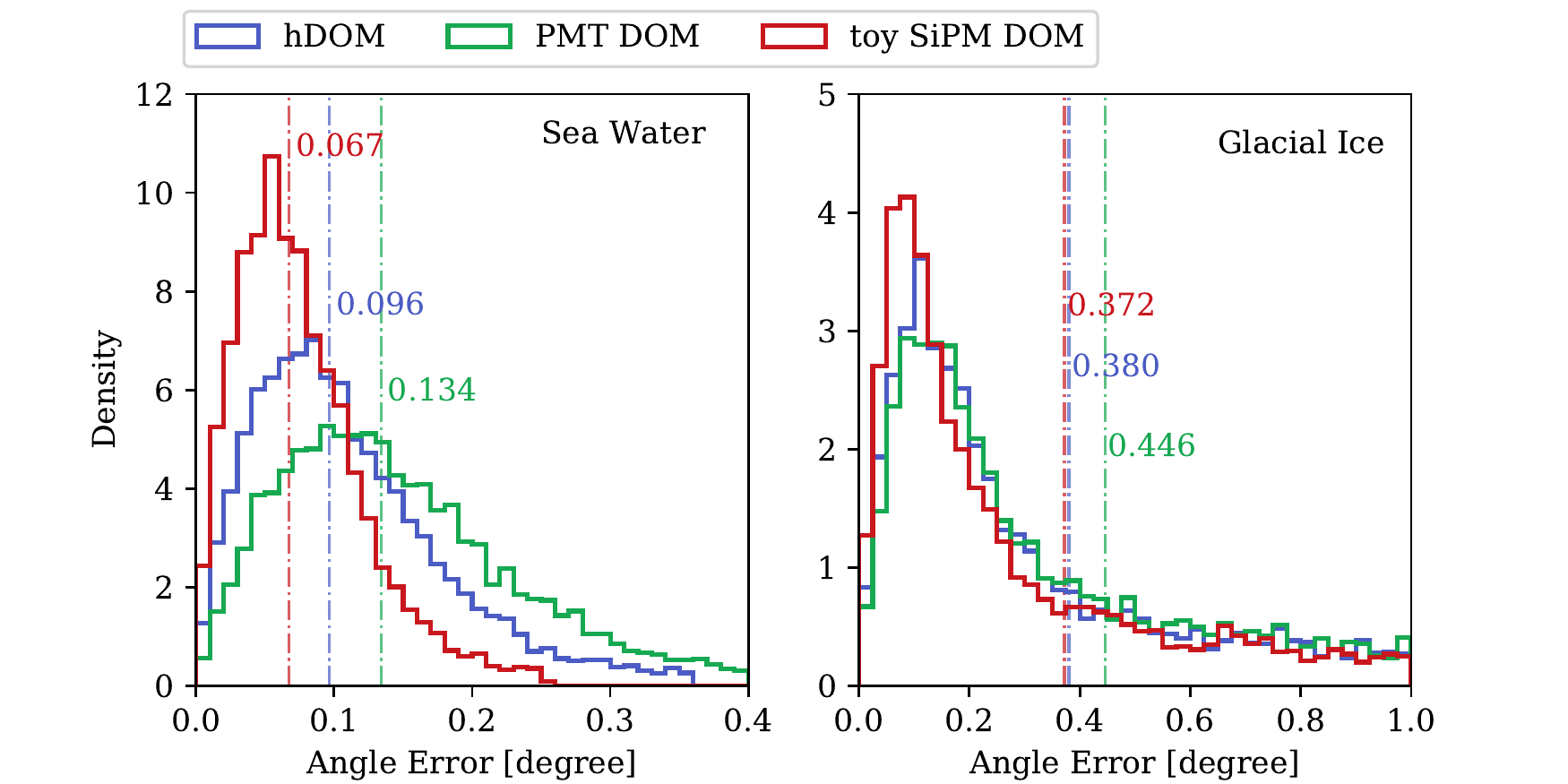}
    \caption{\textbf{Angular error distribution in reconstruction.} \textbf{Left:} Antares sea water, \textbf{right: } IceCube glacial ice. Vertical lines and numbers show the median of angular error in reconstruction.}
    \label{fig: Angular Resolution}
\end{figure}

The angular resolution of hDOM can be boosted by combining the SiPM hits that have a fast time response. As can be seen in the left panel of figure \ref{fig: Angular Resolution}, the angular resolution, represented by the median angular error in reconstruction, of PMT DOM is $0.134 ^\circ$, toy SiPM DOM is $0.067 ^\circ$ and hDOM is $0.096 ^\circ$ in the sea water medium. The toy SiPM DOM has a 100\% improvement in angular resolution compare to the benchmark PMT DOM, clearly indicating the importance of timing ability. The hDOM has a 40\% improvement over the PMT DOM, which is benefited by the additional 20\% hits received by SiPMs. This is can be explained by that the MPE reconstruction method is mostly influenced by the first photon hits information, thus strongly relies on the time accuracy of first hits. 

The high timing accuracy of SiPM has less effect for photons propagating in the ice medium. As shown in the right panel of figure \ref{fig: Angular Resolution}, we can easily observe that the median angular error of benchmark PMT DOM configuration is 2.3 times larger than it in sea water. The angular resolution of toy SiPM DOM and hDOM has 20\% and 17\% gain over the benchmark PMT DOM separately. This is caused by the strong scattering nature of the ice medium. When the distance between track and hDOM is $20 ~\mathrm{m}$, the mean scatter times is 23 for ice medium and 1.8 for sea water medium. In such a situation, the time uncertainty is dominated by the scattering effect.

\section{Summary and Outlook}

We have explored preliminarily an idea of combining SiPMs and PMTs in one hybrid DOM (hDOM) which will increase both photon collection area and pointing capability of a next generation neutrino telescope. A Geant4 simulation and reconstruction pipeline have been set up to analyze 1 TeV muon track events in the neutrino telescope with different DOM configurations: PMT-only DOM for benchmark, toy SiPM DOM to compare with PMT-only DOM, and a more realistic hDOM.

We found that the toy SiPM arrays and hDOM arrays have 100\% and 40\% improvement in angular resolution over the benchmark PMT-only DOM respectively. This proves the benefit of photon sensors’ precision timing in the pointing capability of a neutrino telescope. Such conceptual hDOM configuration could be further optimized to fully captivate the benefits of SiPMs. This fast response timing features of SiPMs, however, does not take significant effect in the glacial ice medium due to the strong scattering effect.

In the future, we plan to simulate higher energy muon events by applying state-of-art simulation tools such as CORSIKA-8 \cite{CORSIKA8} to investigate the impact from substantial stochastic losses, and also with NVIDIA ray-tracing engine OptiX \cite{Blyth:2019yrd} to speed up photon propagation efficiency. The reconstruction method used can be further refined with improved PDF and better characterized hit information. The dark noise of both PMTs and SiPMs can affect the reconstruction and will be studied in detail in the near future. The back-end data read-out system of such an hDOM could be a challenge, but with the rapid progress of modern electronic technologies, it can become feasible sooner than expected.

\bibliographystyle{ICRC}
\bibliography{references}

%
 
\end{document}